\documentstyle[11pt,paspconf,epsf]{article}

\begin{document}

\title{Simulating X-ray Clusters with Adaptive Mesh Refinement}

\author{Greg L. Bryan}
\affil{Physics Department, MIT, Cambridge, MA 02139}

\author{Michael L. Norman}
\affil{National Center for Supercomputing Applications; and \\
Astronomy Department, University of Illinois at
Urbana-Champaign, Urbana, IL 61801
}


\begin{abstract}

Gravitational instabilities naturally give rise to multi-scale
structure, which is difficult for traditional Eulerian hydrodynamic
methods to accurately evolve.  This can be circumvented by adaptively
adding resolution (in the form of multiple levels of finer meshes)
to relatively small volumes as required.  We describe an application
of this adaptive mesh refinement (AMR) technique to
cosmology, focusing on the formation and evolution of X-ray clusters.
A set of simulations are performed on a single cluster, varying
the initial resolution and refinement criteria.  We find that although
new, small scale structure continues to appear as the resolution is
increased,
bulk properties and radial profiles appear to converge at an effective
resolution of $8192^3$. We find good agreement with the ``universal''
dark matter profile of Navarro, Frenk \& White (1995).

\end{abstract}



\section{Introduction}
\label{sec:introduction}

Due to their high luminosity and relative simplicity, X-ray clusters
provide one of the most precise measurements
of the amplitude of mass fluctuations in our universe.  Combined with
the observed anisotropy of the cosmic background radiation, it serves
as a key constraint on cosmological models (\cite{hen92}; \cite{eke96}).
There are, however, a number of unresolved difficulties in our
understanding of clusters.  These include the refusal of clusters
to agree with some analytic scaling laws (\cite{edg91a}), a result
which adiabatic simulations seem unable to explain (\cite{nav95}).
Also, the apparent decrease in the number of X-ray clusters at high
redshift (\cite{cas93}; \cite{bow94}) is unexpected in the 
context of many popular models as well as being discrepant with
optical observations of rich, distant clusters (\cite{cou91};
\cite{pos93}).  Having fixed the number density of clusters at $z=0$,
the rate of evolution is a strong indicator of cosmology, especially
with regard to the value of $\Omega$, thus it is important to
better understand the structure and formation of X-ray clusters.

While N-body studies provide much useful information (\cite{col96}),
clusters are observationally identified either through their galaxies
or by X-ray emission from a hot gas component.  Since galaxies are very
difficult to model correctly, and do not provide as straightforward a
tracer of clusters as X-ray observations, we turn to the baryonic gas.
Most studies of individual X-ray clusters incorporating hydrodynamics
have employed Lagrangian, particle-based methods (\cite{evr90};
\cite{kat93}).  Although these Smoothed Particle
Hydrodynamics (SPH) methods provide excellent spatial resolution when
combined with a suitable gravity solver, their shock-capturing
capabilities are not as good as modern Eulerian methods.  However,
most cosmological Eulerian codes are hampered by a fixed grid and so
provide good resolution in low-density regions, but poor resolution in
high-density regions, such as the centers of X-ray clusters
(\cite{kan94b}).


Here, we present first results from a new method which is designed to
provide adaptive resolution combined with a shock-capturing Eulerian
hydrodynamics scheme.
This Adaptive Mesh Refinement (AMR) technique
provides high resolution within small regions, the location of which
are controlled automatically (\cite{ber89}).
We use the piecewise parabolic method, adapted to cosmology
(\cite{bry95}), for the baryons, particles for dark matter and a
high-resolution gravity solver;
however, due to space constraints, we
defer discussion of the methodology to a 
future paper.






\section{Results}
\label{sec:amr_results}

We have simulated the formation of an adiabatic X-ray cluster in an
$\Omega=1$ universe.  The initial spectrum of density fluctuations is
CDM-like with a shape parameter of $\Gamma = 0.25$ (\cite{efs92b}); the
cluster itself is a constrained 3-$\sigma$ fluctuation at the center,
for a Gaussian filter of 10 Mpc.  We use a Hubble
constant of 50 km/s/Mpc and a baryon fraction of 10\%.  This cluster
is the subject of a comparison project between twelve different
simulation methods, the results of which will be presented in an
upcoming paper (\cite{fre96}).

The simulation was initialized with two grids.  The first is the root
grid covering the entire 64 Mpc$^3$ domain with $64^3$ cells.  The second
grid is also $64^3$ cells but is only 32 Mpc on a side and is centered
on the cluster.  Thus, over the region that forms the cluster, we have
an initial cell size of 500 kpc leading to an approximate mass
resolution of $8.7 \times 10^8 M_\odot$ ($7.8 \times 10^9 M_\odot$)
for the baryons (dark matter).  We adopt a refinement mass for the
baryons of $4 M_{initial} \approx 3.5 \times 10^9 M_\odot$ (i.e. if
the mass in any cell exceeds this value a finer mesh is created), but
only allow refined grids within a box $25.6$ Mpc on a side, centered
on the cluster center since we are uninterested in objects outside
this volume.  Some objects will collapse outside this region and then
move inside; these halos will not be properly modelled as high
resolution is required throughout an object's evolution
(\cite{ann96}).  We will focus mostly on the properties of the central
cluster, which collapsed entirely within the refined region.  We have
also run a set of AMR simulations for the same cluster with lower mass
resolution and initial power in order to examine numerical convergence.

\begin{figure}
\epsfysize=7.4in
\centerline{\epsfbox{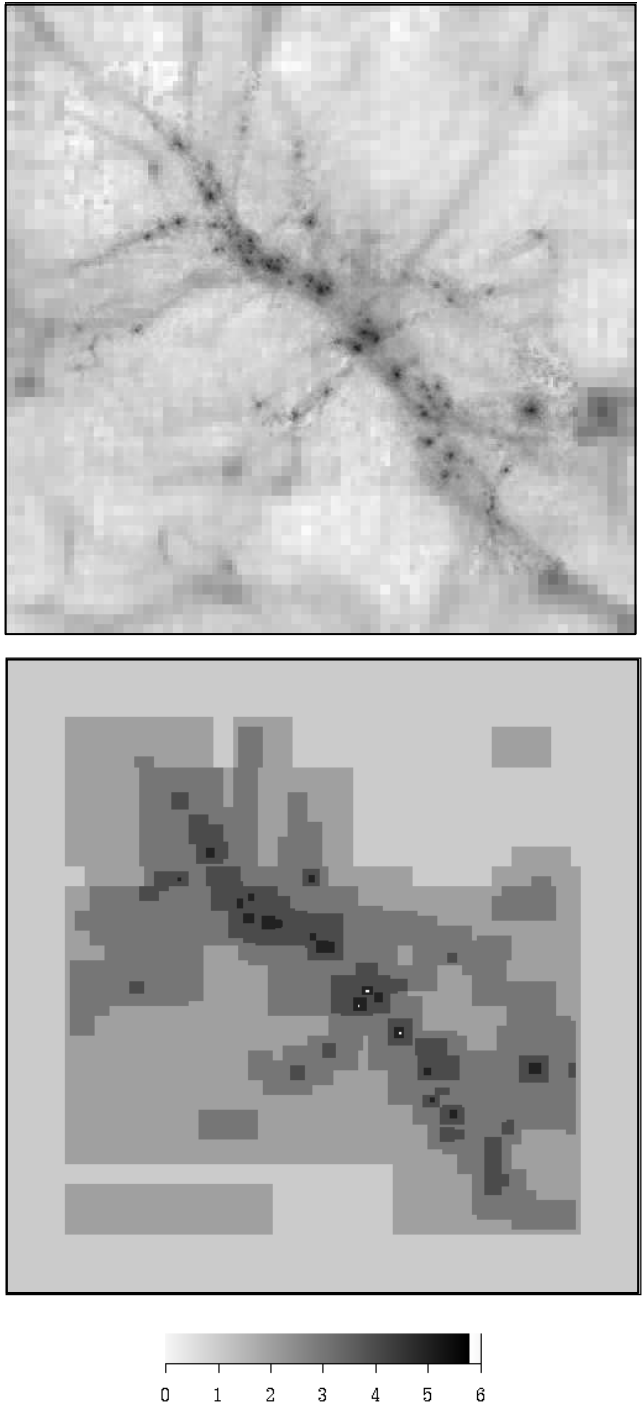}}
\caption{
The logarithm of the dark matter surface density (top) and the
projected grids (bottom), colour coded by level at $z=2$.  In order to
increase the contrast of the three small level 6 grids, they have been
coloured white.  The figures are 32 Mpc on a side.
}
\label{fig:level_example}
\end{figure}

In Figure~\ref{fig:level_example}, we show a typical example of the
grid layout in this simulation.  The top panel depicts the dark matter
distribution in order to show the collapsed structure.  A projection
of the level hierarchy in shown below, with grids colour-coded by
level.  We do not show the full three-dimensional layout since, with
about 400 grids, this would be too complicated to extract much useful
information; however, the grid `shadows' do demonstrate that the grid
structure mirrors the mass morphology.  The range of grid sizes and
shapes is diverse, but most tend to be in the range of 10--60 zones per
edge and largely rectangular.  The images are 32 Mpc on a side; note
the unrefined region around the boundary of the figure.  A
side effect of the quasi-Lagrangian refinement criterion coupled
with the varying cell size is that the {\it rms} density fluctuations
are roughly constant (on a log basis) throughout the hierarchy.
This means that the gravitational force errors due to shot noise
from the finite number of particles is roughly independant of density,
rather than rising sharply in low density regions as in traditional
high-resolution gravity schemes, such as P$^3$M and tree-codes. 


\begin{figure}
\epsfysize=7.1in
\centerline{\epsfbox{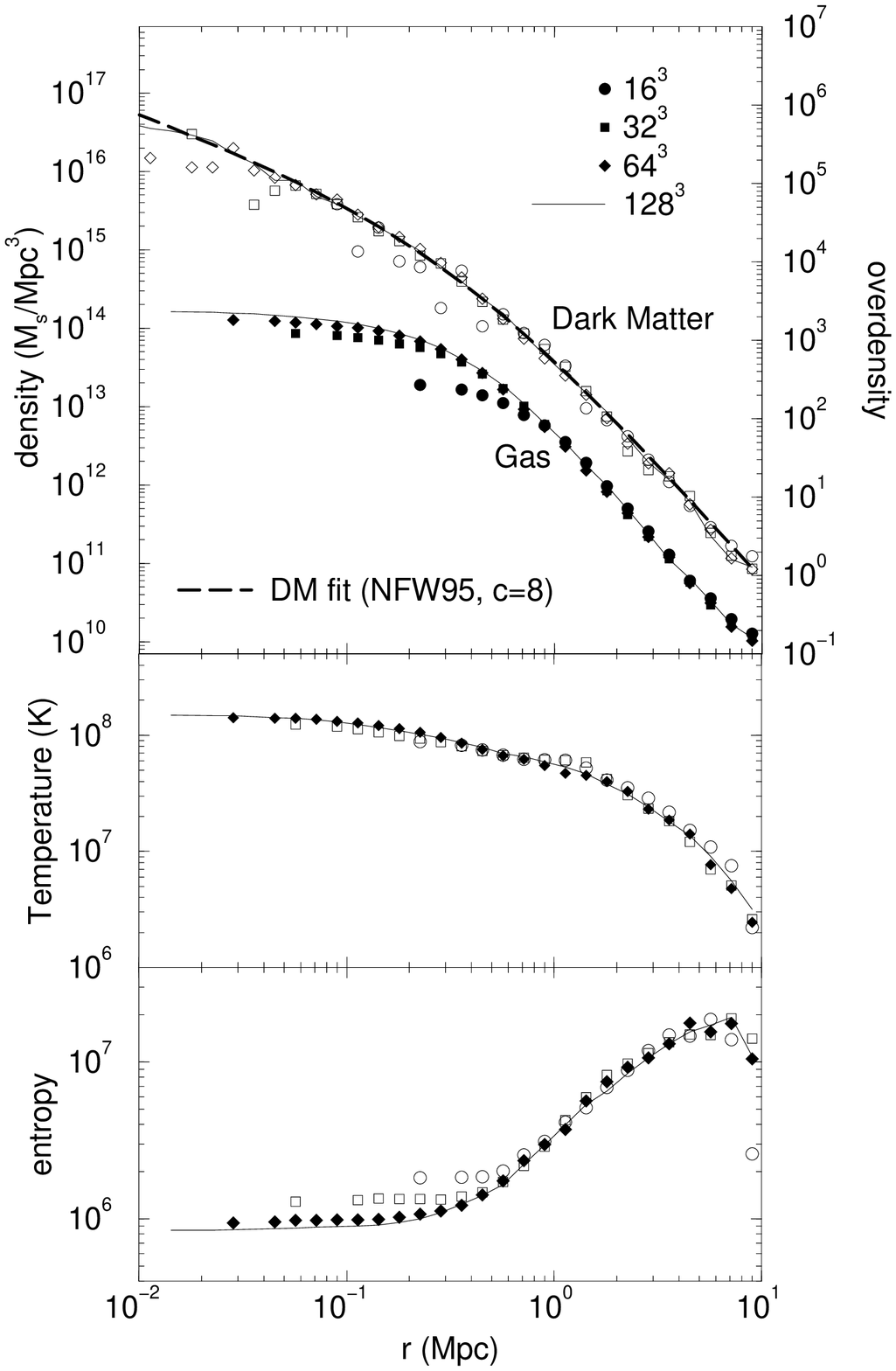}}
\caption{
From top to bottom: i) Dark matter (top curve) and baryonic (bottom
curve) radial density profiles, ii) temperature profile, iii) entropy
profile.  Four different runs are shown with varying initial grid
sizes (which also roughly indiciates the mass resolution):
$16^3$, $32^3$, $64^3$ and the effective $128^3$ run.  The solid
dashed line overlayed the dark matter profile is the fit from
equation~\protect\ref{eq:chapter8_nfw_profile}. 
}
\label{fig:amr_profiles}
\end{figure}

Figure~\ref{fig:amr_profiles} shows the baryonic and dark
matter density profiles for this cluster.  In order to gauge the
convergence, results from three other AMR runs are also plotted.
These runs had smaller initial grids ($16^3$, $32^3$ and $64^3$) and
therefore poorer mass resolution and less initial power.  We also plot
the profile from Navarro et al. (1996) for the dark
matter,
\begin{equation}
\frac{\rho(r)}{\rho_{crit}} = \frac{\delta_0}{(r/r_s)(1 + r/r_s)^2},
\label{eq:chapter8_nfw_profile}
\end{equation}
where $r_s = r_{vir}/c$ and $\delta_0$ is set by the requirement that
the mean density within the virial radius be 178 times the critical
density.  Setting the parameter $c=8$ produces a remarkably good fit
over three orders of magnitude.  


The gas density profile levels off at a few hundred kpc, around the knee
in the dark matter profile.  Lower resolution results exhibit
systematically lower densities, and although we have not converged,
the difference between the $64^3$ initial grid and the
higher-resolution run is slight.  We remind the reader that
this simulation does not include radiative cooling which would
significantly affect the dynamics and structure of the inner few
hundred kpc.
The turnover in density agrees with that seen in entropy, shown in the
same figure.  There appears to be a cutoff in the
entropy distribution, the cause of which is not currently understood.

\section{Conclusion}

The AMR algorithm is complementary to other simulation techniques,
such as SPH and Eulerian single grid method.  The advantage of AMR is
that it provides higher resolution than Eulerian methods and better
shock capturing features than SPH codes.  Further, it is more flexible
than Lagrangian codes because we can control where the resolution is
placed by changing the refinement criterion.  Also, since each level
advances with its own timestep the
entire computation does not have to proceed at the speed of its
slowest component (some SPH codes also share this feature).  
The primary disadvantage is that the scheme is somewhat more
complicated to code and modify; this implementation uses a combination
of C++ to handle the dynamic grid hierarchy and FORTRAN 77 for
computationally intensive tasks.  

Here we have demonstrated that AMR can model an X-ray cluster with many
of the same desirable characteristics of the single-grid code but with
much higher resolution.  The efficiency of the AMR method over a
single grid for this simulation is quite high: a factor of 4000 in
memory and 20 000 in CPU, although, of course, the
resulting solution is not as good in low density regions.


\acknowledgments

We acknowledge useful discussions with Henry Neeman and Edmund
Bertschinger.  This work is done under the auspices of the Grand
Challenge Cosmology Consortium and supported in part by NSF grants
ASC-9318185 and NASA Long Term Astrophysics grant NAGW-3152.


\end{document}